\documentstyle[epsf,emulateapj]{article}

\newcommand{\LCDM}{$\Lambda$CDM\ }
\newcommand{\HH}{H$_2$\ }

\newcommand{\NHI}{N_{\rm HI}}

\newcommand{\cmc}{\,{\rm cm$^{-3}$}\,}

\newcommand{\kmsmpc}{\,{\rm km\,s$^{-1}$\,Mpc$^{-1}$}\,}
\newcommand{\ev}{\,{\rm eV\ }}
\newcommand{\kel}{\,{\rm K\ }}
\newcommand{\etal}{{ et~al.~}}
\newcommand{\fluxunit}{\,{\rm erg\,s$^{-1}$\,cm$^{-2}$\,Hz$^{-1}$\,}}
\newcommand{\Ms}{M_\odot}
\newcommand{\MsMpc}{M_\odot\,\hbox{\rm Mpc$^{-3}$}}
\newcommand{\specunit}{\,{\rm erg\,s$^{-1}$\,cm$^{-2}$\,Hz$^{-1}$\,sr$^{-1}$\,}}

\begin{document}


\title{Simulations of Pregalactic Structure Formation with Radiative
Feedback}

\author{Marie E. Machacek}
\affil{Physics Department, Northeastern University, Boston, MA 02115}

\author{Greg L. Bryan\altaffilmark{1}}
\affil{Physics Department, Massachusetts Institute of Technology,
                Cambridge, MA 02139}

\author{Tom Abel}
\affil{Harvard Smithsonian Center for Astrophysics, 60 Garden Street,
                Cambridge, MA 02138}

\altaffiltext{1}{Hubble Fellow}


\begin{abstract}

We present results from three-dimensional hydrodynamic simulations of
the high redshift collapse of pregalactic clouds including feedback
effects from a soft \HH photodissociating UV radiation field.  The
simulations use an Eulerian adaptive mesh refinement technique to
follow the nonequilibrium chemistry of nine chemical species with
cosmological initial conditions drawn from a popular
$\Lambda$-dominated cold dark matter model.  The results confirm that
the soft UV background can delay the cooling and collapse of small
halos ($\sim 10^6 \Ms$).  For reasonable values of the
photo-dissociating flux, the \HH fraction is in equilibrium throughout
most of the objects we simulate.  We determine the mass threshold for
collapse for a range of soft-UV fluxes and also derive a simple
analytic expression.  Continuing the simulations beyond the point of
initial collapse demonstrates that the fraction of gas which can cool
depends mostly on the virial mass of the halo and the amount of
soft-UV flux, with remarkably little scatter.  We parameterize this
relation, for use in semi-analytic models.

\end{abstract}



\section{Introduction}
\label{sec:introduction}

A critical epoch in the evolution of the universe is the
the post-recombination period during which the 
first stars form, affect their environment and help determine how the 
intergalactic medium (IGM) ultimately becomes reionized. Recent observations
are beginning to constrain models of this epoch.  The observation of quasars 
at redshifts as high as  $z=5.8$ (Fan \etal 2000) with no accompanying 
Gunn-Peterson absorption trough in their spectra indicates that the onset of 
reionization occurs quite early ($z \ge 7$) and so the first generation of 
stars forms even earlier. Observations of quasar spectra 
(Ellison \etal 1999, Ellison \etal 2000) suggest that metallicities possibly 
as high as $\log CIV/HI = -2.6$ may be present even in the low 
column density $\log \NHI < 14.5$ Lyman alpha forest. This too may be 
indicative of an active period of pre enrichment of the IGM by an early 
generation of stars (Ostriker \& Gnedin 1996).  Future multiwavelength 
observations may be sensitive probes of this star formation epoch and the 
subsequent onset of reionization. A partial list of proposed observations 
include direct imaging of quasars and star clusters at $z > 10$ by the Next 
Generation Space Telescope (Haiman \& Loeb 1999, Barkana \& Loeb 2000), 
absorption measurements in the 21 cm line using the Square Kilometer Array 
to identify the first epoch of massive star formation (Tozzi \etal 1999), 
and searches for secondary anisotropies in the Cosmic Microwave
Background by the next generation of millimeter telescopes such as
ALMA to constrain properties of the reionization epoch (Benson \etal
2000). Thus it is important to understand what our best cosmological
models predict for the evolution of these early star forming
structures and what governs the amount of gas that can ultimately
condense into stars.

It has long been understood that a high redshift population of stars could 
form naturally in hierarchical models of structure formation.  In these 
models small masses collapse first and later merge into larger structures.  
The ability of these objects to form stars depends on the ability of gas 
within these objects to cool radiatively, reduce pressure support, and 
condense.  However, the first structures to become self-gravitating 
in these models would have temperatures $T < 10^4 \kel$, too cool for 
HI line cooling to be effective. The only other coolant available to gas 
with a primordial composition is molecular hydrogen. Therefore it is the 
complex chemistry of \HH in these objects that controls whether the gas can 
condense to sufficiently high densities for star formation to take place.  

Early analytical models of the collapse of the first small scale structures
studied the effect of \HH cooling within highly idealized 
cloud geometries, typically a homogeneous gas cloud radiatively cooling while 
undergoing a uniform free fall collapse (Saslow \& Zipoy 1967, Peebles \& 
Dicke 1968, Hirasawa 1969, Matsudo \etal 1969, Hutchins 1976, Silk 1983, 
Palla \etal 1983, Lepp \& Shull 1984, Lahov 1986).  1-D simulations sought
to improve upon these analytical estimates, but still had to impose the 
shape of the initial density profiles and mass of the virialized object as
initial conditions within an assumed cloud geometry (Haiman, Thoul \& Loeb 
1996, Tegmark  \etal 1997, Omukai \& Nishi 1998, Nishi \etal 1998, 
Nakamura \& Umemura 1999). Haiman, Thoul \&  Loeb (1996) followed the 
dynamics of gas and cold dark matter and the \HH chemistry in a single, 
spherical density peak and showed the importance of shell crossing 
between the gas and dark matter in determining the fraction of mass within 
an object that could collapse and the minimum mass for collapse.  In the 
same spirit Tegmark \etal (1997) used top hat density profiles and a 
simplified collapse criterion, i.e. that the cooling time be less than the 
Hubble time for the object right after virialization, to obtain rough 
estimates of the critical \HH mass fraction needed for cooling and the 
minimum mass (as a function of virialization redshift) that could produce 
it.  However, the minimum mass to collapse determined in this way is 
strongly dependent on the \HH cooling function used. Abel \etal (1998) and 
Fuller \& Couchman (2000) recalculated this minimum mass for top hat density 
profiles using the Tegmark \etal (1997) collapse criterion with updated \HH 
cooling functions (Lepp \& Shull 1983 and Galli \& Palla 1998, respectively) 
and found that the minimum mass able to collapse (at virialization redshifts 
$\sim 30$) was reduced by as much as an order of magnitude.  

Molecular hydrogen is fragile and easily photodissociated by photons
whose energies lie below $13.6 \ev$ (Stecker \& Williams 1967).  Thus
even before reionization, radiation from the first stars could affect
\HH abundances at large distances (Dekel \& Rees 1987).  Haiman, Rees,
\& Loeb (1997) quantified this negative radiative feedback in more
detail.  They studied the collapse of homogeneous, spherical gas
clouds in the presence of a soft UV photodissociating flux assuming
both ionization and chemical equilibrium within the cloud to determine
which masses could still collapse. Haiman, Abel \& Rees (1999)
improved on that calculation by solving spherically symmetric
radiative transfer coupled with the time dependent rate equations for
the \HH chemistry across truncated isothermal spheres. They also
modified the collapse criterion such that a structure was said to be
able to collapse and form stars if at any time during the object's
history the \HH cooling time in the core became less than the time
elapsed since the core's formation.  This allowed the fraction of \HH
abundance required for cooling and collapse of the core to be lower in
certain cases than previous estimates of the ``critical'' \HH fraction
based on the Tegmark \etal (1997) collapse criterion.  Analytical
models that include a first population of OB-like stars have been used
to probe the suppression of star formation elsewhere within the host
cloud due to the destruction of \HH coolant by the direct radiation
from the first stars (Omukai \& Nishi 1999; Glover \& Brand 2000).

The major drawback affecting all of the above studies whether with 
or without radiative feedback is their one dimensional nature. Hierarchical 
structure formation that predicts the growth of structure in and along 
filaments is inherently three dimensional. However, fully three dimensional 
simulations are computationally challenging due to the need for large dynamic 
range to model simultaneously the internal structure of the first star 
forming structures and the effects of gravitational tidal forces and merging 
on their growth. Such 3-D simulations have only recently become 
computationally feasible. Gnedin \& Ostriker (1997) simulated the thermal 
history of the intergalactic medium and the population of first objects to 
collapse but had insufficient resolution to probe inside these first 
structures.  Abel \etal (1998) used a static nested grid approach 
and constrained realizations of $3$ -- $4 \sigma$ primordial fluctuations in 
a cluster normalized SCDM model to study when these structures were 
able to cool, but were also limited by spatial resolution 
($\sim 1$~kpc) from probing deeply within the peaks.  Subsequent higher 
resolution 3-D simulations focused on the properties of a single 
collapsing cloud.  Bromm, Coppi, \& Larson (1999) studied the fragmentation 
of a $2 \times 10^6 \Ms$ rotating disk under highly idealized initial 
conditions using a TREESPH code.  Fuller \& Couchman (2000) used an
adaptive smoothing-kernal SPH coupled to a P$^3$M gravity solver in a  
simulation box of comoving length $25 h^{-1}$~kpc to investigate the 
properties of the first peak to collapse. However, because of the small box 
size they needed to impose an overdense cold dark matter cosmology to model 
the formation of the cloud within its host filament. The highest spatial and 
mass resolution self-consistent cosmological 3-D simulations to date of the 
collapse and fragmentation of primordial clouds (Abel, Bryan, \& Norman 1998,
 2000; Norman, Abel, \& Bryan 2000) use Eulerian adaptive mesh refinement 
(AMR) techniques to achieve extremely large dynamic range 
(currently $ \ge 3 \times 10^7$). They start from cosmological initial
conditions in a cluster normalized SCDM model to study the detailed internal 
evolution of a single, collapsing cloud in interaction with its environment 
in a box with comoving length $L=128$~kpc. 

In this paper we present the results of fully 3-D Eulerian AMR    
simulations of the formation and collapse of primordial pregalactic structure
in a realistic \LCDM cosmology.  This study expands upon previous work in 
several important ways. First, we use a larger simulation volume ($1$~Mpc$^3$ 
comoving) to model more accurately the gravitational tidal forces and 
effects of merging on the evolving structures.  Second, we introduce a 
simple model for radiative feedback. We analyze the collapse properties of 
individual peaks with and without the presence of this soft UV background and 
compare our results with previous 1-D studies.  Finally, we use the 
population of pregalactic objects evolving together in the most active 
structure forming region of our simulation to develop statistics on the 
fraction of gas that is able to cool due to the presence of \HH and the 
fraction of gas that is both cold and dense enough to be available for star 
formation as functions of cloud mass and the average soft UV radiation field 
it experiences. These statistics represent the key new result of this paper. 
The amount of gas available for star formation in pregalactic objects is a 
critical input parameter into recent semi-analytical attempts to model 
stellar feedback from the first stars on the intergalactic medium, 
subsequent structure formation, and the nature of reionization 
(Gnedin \& Ostriker 1997, Ciardi \etal 2000, Ferrara \etal 2000, 
Ciardi, Ferrara, \& Abel 2000).    

While we focus on the negative effect of radiation in this paper, it
should be noted that positive feedback is possible if the first
objects also emit high energy X-ray photons (Haiman, Abel \& Rees
1999; Oh 2000).  We will address this point in future work.

This paper is organized in the following way. In Section \S\ref{sec:setup} we 
describe the set-up of our simulations including simulation parameters and
technique, cosmological model, \HH chemistry, and radiative feedback.    
In Section \S\ref{sec:dataset} we present the characteristics of the 
simulated data set of pregalactic objects identified in our simulations and  
used for further analysis. In Section \S\ref{sec:stats} we use the data set 
as a whole to develop fitting formulaes for the fraction of gas that can 
cool, become dense and thus be available for star formation with and without 
the presence of a background \HH photodissociating flux. In Section 
\S\ref{sec:profile} we use radial profiles to show the internal collapse 
characteristics of these structures. We consider both the time evolution of 
a single peak for a fixed background \HH photodissociating flux and also how 
the internal properties of a given cloud change at fixed redshift when the 
level of \HH photodissociating flux is varied. Armed with these examples of 
the internal evolution and structure of the collapsing clouds, we discuss in 
Section \S\ref{sec:discussion} the
neglect of self-shielding and H$^-$ photodetachment in our simulations and 
then use a simple analytical argument to illuminate the essential physics of 
the collapse.  We summarize our results in Section \S\ref{sec:summary}.


\section{The Simulations}
\label{sec:setup}

A self-consistent three dimensional cosmological simulation of the collapse
of low mass ($10^5$ -- $10^7 \Ms$) protogalaxies at high redshift 
($30 > z > 19$) requires both large enough simulation
volume to model the gravitational tidal forces at work during the formation
epoch and high spatial (and mass) resolution to probe within the small 
collapsing clouds.  One technique that has recently been able to achieve 
the large dynamic range necessary to shed light on the 
collapse and fragmentation of these first star producing structures is 
adaptive mesh refinement (AMR).  The three dimensional AMR algorithm used in 
our simulations is described more fully elsewhere (Bryan 1999; Bryan \& 
Norman 1997, 1999; Norman \& Bryan 1998).  Briefly, the code uses an 
adaptive hierarchy of rectangular grid patches at various levels of 
resolution very similar to the algorithm described by Berger \& Collela 
(1989). Each grid patch covers some region within its parent grid needing 
additional refinement and may itself become a parent grid to an even higher 
resolution child grid. We take the ratio of parent to child grid mesh 
spacing to be two. The dark matter is followed using methods similar to 
those presented by Couchman (1991) and the gas hydrodynamics uses the 
artificial viscosity method of Stone \& Norman (1992). 

We use a flat, low matter density \LCDM model for our simulations with 
$\Omega_0=0.4$, $\Omega_b=0.05$, $\Omega_\Lambda=0.6$, $h=0.65$, 
$\sigma_8=0.8$, and $n=1$ where $\Omega_0$, $\Omega_b$, 
and $\Omega_\Lambda$ are the fraction of the critical energy density carried 
in nonrelativistic matter, baryons, and vacuum energy, respectively, $h$ is 
the dimensionless Hubble parameter in units of $100$ \kmsmpc, $\sigma_{8}$ 
is the density fluctuation normalization in a sphere of radius $8h^{-1}$ Mpc,
and $n$ is the slope of the primordial density perturbation power spectrum.  
The parameters are chosen to provide good consistency with observation.  The 
fluctuation normalization is consistent with the CMB quadrupole as measured
by COBE (Bunn \& White 1997) and also with observations of the number density
of galaxy clusters (White, Efstathiou \& Frenk 1993; Bond \& Myers 1996). 
$\Omega_b h^{2}$ is consistent with big bang nucleosynthesis and the measured
abundance of primordial deuterium (Copi, Schramm, \& Turner 1995; Burles \& 
Tytler 1998), and $\Omega_\Lambda$ is consistent with the upper limit 
($\Omega_\Lambda < 0.7$) of Maoz  \& Rix (1993) and the best fit parameters
of Ostriker \& Steinhardt (1995).     

The simulations are initialized at $z=99$ with density perturbations
generated for the above \LCDM model using the Eisenstein \& Hu (1998)
transfer functions in a comoving simulation volume of $1$~Mpc$^3$.
This simulation volume is large enough to ensure that the fundamental
mode in the box remains linear at the lowest redshifts considered
here.  A low resolution run is used first to identify a region of
active structure formation.  In order to achieve high mass resolution
in the region of interest the simulation is reinitialized with
multiple static refinement levels surrounding the chosen region such
that each successive static parent grid contains the lagrangian volume
of its child grid.  This results in a mass resolution in the initial
conditions for the region of interest of $4.78 (38.25) \Ms$ for the
gas (dark matter), respectively.  In addition, that region is allowed
to dynamically refine so that the local Jeans length is resolved at
all times by at least four grid zones and no grid cell contains more
than four times the initial gas mass element ($4.78 \Ms$) or ten
times the initial dark matter element ($382.5 \Ms$).  We limit the
total refinement to $14$ levels within a $64^3$ top grid which results
in a maximum dynamic range of $1,048,576$.  The comoving spatial
resolution of $0.95$~pc at maximum refinement translates into a
physical spatial resolution of $0.03$~pc ($0.05$~pc) at $z=30$
($z=19$), respectively.  We call a peak ``collapsed'' when it reaches
maximal refinement. Once this occurs, we still wish to follow the
subsequent evolution of structure in the region, but are uninterested
in the details of the collapsed peak.  In order to prevent the peak
from collapsing further (which could cause numerical instabilities),
we introduce a form of artificial pressure support.  This is done by
defining, for each cell, an effective pressure which is the greater of
the thermal pressure and $K G \rho_b^2 \Delta x_f^2 / \mu$, where
$\rho_b$ is the baryon density in the cell, $\Delta x_f$ is the cell
width on the finest level and $\mu=1.22 m_H$ is the usual mean mass
per particle.  The dimensionless constant $K$ is set to 100, a value
which spreads the mass over a spherical region with a radius of
several cells.  The form of this expression is chosen by matching the
thermal energy and gravitational self-energy of the gas; notice also
that it results in a polytropic equation of state.  We also smooth the
gravitational potential by $4 \Delta x_f$ which helps to reduce the
amplitude of the artificial pressure support required.  Although
introducing artificial pressure support causes us to lose information
about the detailed morphology of the inner $1$ -- $2$ pc ($\sim 1$ \%
of the virial radius) of the peak after collapse, it should not affect
the determination of cooled gas fractions in Section \S\ref{sec:stats}
since the cooling region lies outside of the collapsed \HH core.

Our data set consists of results from four simulations starting from
identical cosmological initial conditions on the density fields.
Three of the simulations follow the nonequilibrium, time dependent
evolution of nine chemical species (H, H$^+$, He, He$^+$, He$^{++}$,
e$^-$, \HH, H$^+_2$, H$^-$) using the algorithm of Anninos, \etal
(1997) initialized with post-recombination abundances (Anninos \&
Norman 1996).  The reaction network for even this simple chemistry is
complex. Numerous authors have worked to identify the dominant
reactions and calculate their rate coefficients including recently
Haiman, Thoul \& Loeb (1996), Abel \etal (1997) and Galli \& Palla
(1998).  We use primarily the rate coefficients of Abel \etal (1997)
with the exception of rate coefficients for the processes $H + e^-
\rightarrow H^- + \gamma$, $H_2 + e^- \rightarrow 2H + e^-$, and $H^-
+ H^+ \rightarrow H_2^+ + e^-$ which are taken from Shapiro \& Kang
(1987). We have also included the three-body process $2H + H
\rightarrow H_2 + H$ (Orel 1987, Palli \etal 1983), although its
contribution is negligible at the densities considered in this work.
The small post-recombination abundance of \HH in the intergalactic
medium ($\sim 2 \times 10^{-6}$) is too low for cooling to be
effective. However, at the densities and temperatures in the interiors
of collapsing proto-galactic clouds, \HH formation is catalyzed by
H$^-$ ions through the process
\begin{equation}
H + e^- \rightarrow H^- + \gamma
\label{eq:hminusform}
\end{equation}
\begin{equation}
H + H^- \rightarrow H_2 + e^- .
\label{eq:hminus2h2} 
\end{equation}
and can reach abundances high enough to act as an efficient coolant 
allowing the first stars to form.  

Once the first stars form, radiation above $13.6$~eV produced by these stars 
is quickly absorbed by the dense neutral hydrogen surrounding them.  However, 
the intergalactic medium is optically thin to radiation below the HI 
ionization limit.  This radiation escapes to form a soft ultraviolet 
background of radiation throughout neighboring regions with sufficient energy 
to photodissociate \HH by means of the Solomon process
\begin{equation}
H_2 + \gamma \rightarrow H_2^* \rightarrow H + H ,
\label{eq:solomon}
\end{equation}
thus inhibiting continued star formation (Haiman, Abel \& Rees 1999). 
$H_2^*$ is any of $76$ Lyman-Werner resonances in the energy range 
$11.18$ -- $13.6 \ev$.  We use a very simple model to investigate the 
negative feedback effects of this soft UV radiation.  The radiation 
field is assumed to be constant throughout the simulation with mean photon 
energy of $12.87 \ev$. In terms of the constant mean flux $F_{LW}$,
the \HH photodissociation rate coefficient for the Solomon process denoted in
Equation \ref{eq:solomon} is (Abel, \etal 1997)
\begin{equation}
k_{diss} = 1.1 \times 10^8 F_{LW}\,\hbox{\rm s$^{-1}$}.
\label{eq:kdiss}
\end{equation}
We distinguish three of our simulations by different 
levels of this background flux, i.e. $F_{LW}=0$ (first structure 
formation), $10^{-22}$, and  $10^{-21}$ \fluxunit, respectively.
These flux levels correspond to $J_{21}$ normalizations of the 
spectral intensity (Haiman, Abel, \& Rees 1999) of $J_{21} = 0$, $10^{-2.1}$ 
and $10^{-1.1}$ \specunit, respectively. 
We also include results from a fourth simulation that starts from the same 
cosmological initial conditions, but evolves only six species 
(H, H$^+$, He, He$^+$, He$^{++}$, e$^-$) representing the 
extreme case when destruction of molecular hydrogen is complete.

We neglect the time dependent turn on and build-up of the soft
radiative field. Thus the UV radiation experienced by the collapsing
structures may be interpreted as due either to ``direct'', nearby
sources or a more diffuse soft background. We also neglect the effect
of self-shielding by the collapsing cores.
As will be discussed more fully in Section \S\ref{sec:discussion}, this
should have only a small affect on our results.  In any case each of
these neglected effects would tend to reduce the impact of the
photodissociating flux on the collapsing clouds. Thus our results for
the simulations with $F_{LW} = 10^{-22}$ and $10^{-21}$ \fluxunit
should be interpreted as representing the maximal suppression of
structure formation at these flux levels.

We also neglect the photo-detachment of H$^-$ by the UV radiation field.
One might argue that the presence of the radiative background might 
photo-detach H$^-$ before it could catalyze \HH formation by the process 
given in Equation \ref{eq:hminus2h2}. This could hinder the collapse of
clouds by preventing the build-up of coolant in the protogalactic cores. 
While the photons clearly have sufficient energy ($E_\gamma > 0.755 \ev$) to 
photo-detach H$^-$, we will show in Section \S\ref{sec:discussion} that  
H$^-$ photo-detachment is suppressed by at least two orders of magnitude 
relative to \HH formation at the temperatures, densities and UV background
flux levels we consider.


\section{The Sample of Simulated Halos}
\label{sec:dataset}

\begin{figure}
\epsfxsize=5.5in
\centerline{\epsfbox{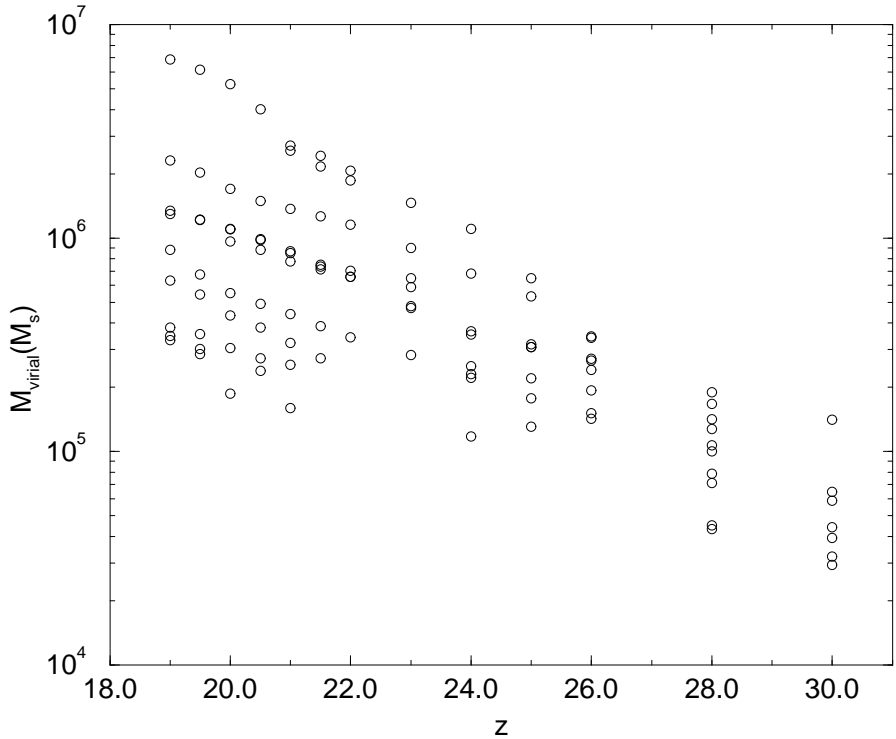}}
\caption{The virial mass as a function of redshift for peaks used in the 
analysis sample for the case with no \HH present. 
 }
\label{fig:mvir_z}
\end{figure}

The locations of high density pregalactic clouds are identified in the 
simulations using the HOP algorithm (Eisenstein \& Hut 1998) acting on the 
dark matter density distribution. We find that the high density peaks form 
within and at the intersection of sheets and filaments as expected in any 
cold dark matter dominated cosmology (Abel, \etal 1998). We choose 
the $8$ --$10$ most massive peaks of the $\sim 200$ identified by HOP at each 
redshift for further analysis.  In Figure \ref{fig:mvir_z} we 
present the mass of these clouds as a function of redshift for 
the simulation without \HH. Since we use identical initial conditions for 
each of the four simulations, this is essentially the 
same set of peaks used for analysis in the simulations with \HH
present and varying levels of the radiative background flux. These peaks 
span a mass range from $3 \times 10^4$ to $7 \times 10^6 \Ms$ and are 
modeled in our simulations by $\sim 1000$ to $125,000$ dark matter 
particles, respectively. The  absence of low mass peaks in the bottom 
portion of Figure \ref{fig:mvir_z} is an artifact
of our selection criterion.  However, the increase in the maximum mass of 
the cloud with decreasing redshift is real and characteristic of the 
``bottom up'' growth of structure in these models. Small clouds form first at 
high redshift, merging to form more massive structures later. There is clear 
evidence for mergers among the clouds we consider.  See, for example, the 
merger of the two most massive peaks in Figure \ref{fig:mvir_z} into a 
single peak between redshifts $z=21$ and $z=20.5$.  However since we do not 
follow all of the low mass clouds in the simulation, we do not construct
complete merger trees for the sample. Thus we caution the reader that all 
the peaks are not statistically independent of each other.  This is 
particularly noticeable for the high mass peaks which are rare in our 
simulation volume and may result in a decrease in the scatter of 
measured properties for the higher mass clouds.

\begin{figure}
\epsfxsize=5.5in
\centerline{\epsfbox{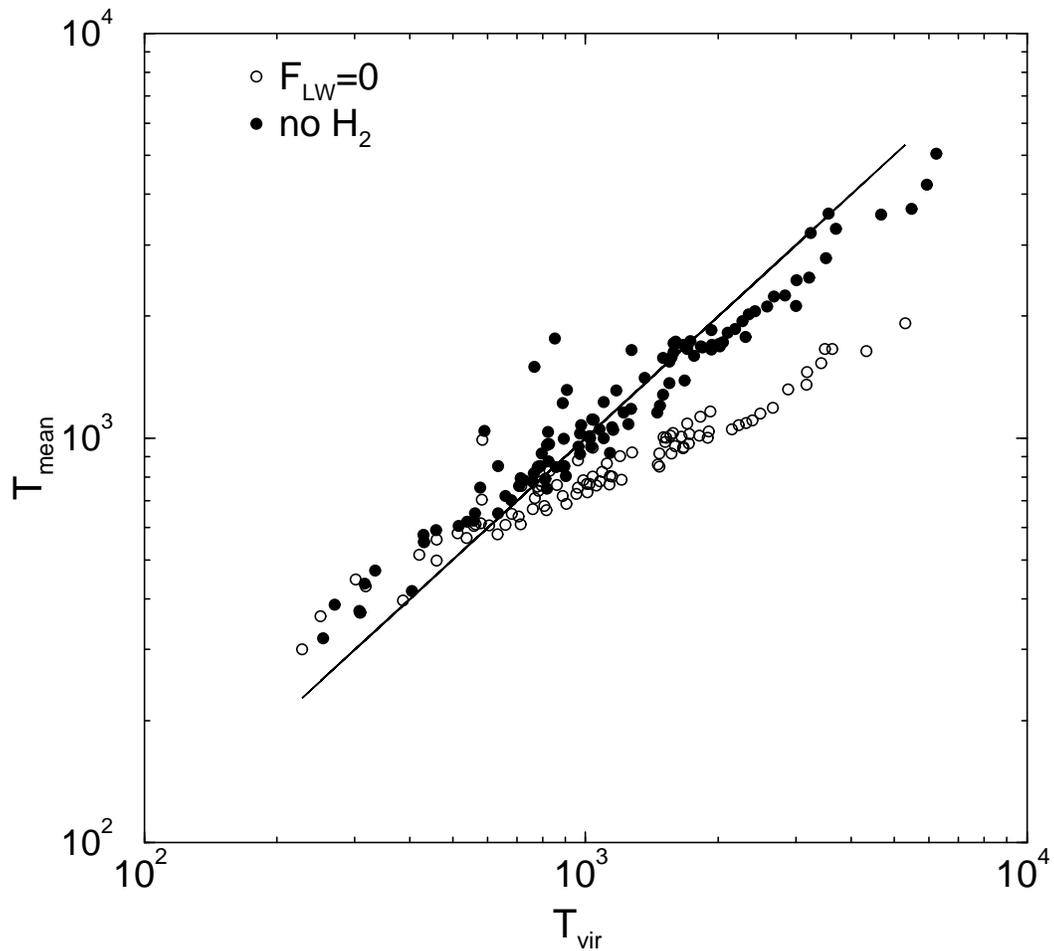}}
\caption{Comparison of the mean gas-mass-weighted temperature $T_{mean}$ for 
the peaks used in our sample with the virial temperature $T_{vir}$ given by 
Equation \protect\ref{eq:mtz} for the case with no \HH cooling
(filled circles) and the case with maximal \HH cooling (open circles). The 
solid line represents the relationship $T_{mean}=T_{vir}$.}
\label{fig:tvir}
\end{figure}

We define the virial radius, $r_{vir}$, as the radius of a sphere within 
which the average density in the cloud exceeds $200$ times the mean matter
density of the universe. The virial mass, $M_{vir}$ is the total mass 
contained within that radius.  Before cooling becomes important the mean
gas-mass-weighted temperature of the cloud obeys a virial relationship 
between temperature, mass and redshift similar to that observed in galaxy
clusters (Bryan \& Norman 1998).  For a neutral mix of primordial gas 
\begin{equation}
T_{vir}(\kel) = 155 f(h^2 \Delta_c E^2)^{1/3}({M_{vir} \over 10^7 
M_\odot})^{2/3} 
\label{eq:mtz}
\end{equation}
where $f=0.8$ is a normalization factor, $h=0.65$ is the dimensionless 
Hubble parameter at $z=0$, $\Delta_c = 200$ is the collapse overdensity and 
\begin{equation}
E^2 = \Omega_0 (1+z)^3+\Omega_\Lambda \approx \Omega_0 (1+z)^3
\end{equation}
for $\Omega_0 = 0.4$ and $\Omega_\Lambda = 0.6$ at redshifts $30 > z > 19$.
In Figure \ref{fig:tvir} we compare the mean gas-mass-weighted 
temperature $T_{mean}$, for the clouds in our sample to the virial 
temperature $T_{vir}$ defined by Equation \ref{eq:mtz} for the case
when no molecular hydrogen is present and for the case $F_{LW}=0$ when 
molecular hydrogen cooling is maximal. All of the
clouds in our sample have virial temperatures 
$T_{vir} < 8000$ -- $10,000 \kel$, the temperature necessary for HI line 
cooling to become important. Peaks analyzed from the simulation without \HH 
can not cool and their mean temperatures are well characterized by the 
virial relation given in Equation \ref{eq:mtz}.  
Once cooling becomes important, the mean gas-mass-weighted temperature of 
the cloud will drop significantly below its virial value as demonstrated in 
Figure \ref{fig:tvir} by peaks from the simulation that includes the \HH 
chemistry. 

In Table \ref{tab:first} we compare the properties of the first protogalaxies
to ``collapse'', i.e reach maximal refinement, for the simulations with \HH
cooling and soft UV background flux levels $F_{LW}=0$, $10^{-22}$, and 
$10^{-21}$ \fluxunit.  Protogalactic clouds in the simulation without \HH
have not collapsed by redshift $z=19$.  The first structures ($F_{LW}=0$) 
in this \LCDM model collapse early ($z \sim 30$) and are small 
($r_{vir} \sim 50$~pc), low mass ($M_{vir} \sim 1.4 \times 10^5 \Ms$) clouds.
Simulations of the minimum collapse mass by Fuller \& Couchman (2000) find 
the minimum mass to collapse at $z=30$ to be $\sim 3 \times 10^5 \Ms$ 
(expected to be robust within about a factor of two) thus consistent with our 
results.  Abel, Bryan, \& Norman (2000) find a $7 \times 10^5 \Ms$ peak 
collapses first (at $z=19$) in their high resolution SCDM simulation. This 
too is consistent with the minimum mass collapse curves of Fuller \& 
Couchman. Thus despite differences in numerical method and cosmology there 
appears to be general agreement that the minimum mass of the first star 
forming structures is $ \sim 10^5 \Ms$.  This is about an order of magnitude 
below the earlier estimates of Tegmark \etal (1997) due primarily to the use 
of improved \HH cooling functions.      

As the amount of the \HH photodissociating flux increases, the mass 
threshold for collapse increases pushing the redshift of collapse to 
lower values.  However, even with flux levels as 
high as $F_{LW}=10^{-21}$~\fluxunit, \HH can survive in the cores of the 
clouds such that objects ($M_{vir} \sim 1.26 \times 10^6 \Ms$, 
$T_{vir} \sim 2262 $\kel) below the mass threshold for HI line cooling are 
still able to collapse. However, the collapse in this case is delayed until
$z=21.5$.  This is consistent with the 1-D simulations of 
Haiman, Abel \& Rees (1999). See, in particular, their Figure 6 for 
$J_{21}=10^{-1.1}$. Furthermore the ability of a cloud to collapse does not 
seem to be strongly dependent on the angular momentum $L$ of its dark matter 
halo, since the threshold objects we found to collapse spanned the full 
range of expected spin parameters $0.02 < \lambda_{dm} < 0.11$ where 
$\lambda = (L|E|^{1/2})/(GM^{5/2})$, $E$ is the total energy, $M$ is the 
mass, and $G$ is the gravitational constant.


In order to check the numerical robustness of our results, we ran
several lower mass resolution simulations ($\Delta M =306 \,(38) \Ms$
for dark matter(gas)) using the same random seed for the initial
conditions and region for dynamical refinement as the high mass
resolution runs presented here.  We also ran several different
realizations of the initial conditions at this lower resolution for
the $F_{LW}=10^{-21}$ \fluxunit case to ensure that there was nothing
exceptional about the particular realization we present.  We found
that, although the properties of the first peak to collapse were very
similar for the two mass resolutions, the higher mass resolution run
delayed the redshift of first collapse to slightly lower values
($\Delta z \sim 1$).  This difference in the redshift at which maximal
refinement is reached, however, is already less than the cosmic
variance ($\Delta z \sim 2$) in this quantity observed between
simulations run with different realizations of the initial conditions.
The mass threshold for collapse was also reduced by $\approx 30$ \% in
the higher mass resolution run compared to the lower resolution
simulation.  This is slightly larger than the $ \pm 20$ \% variation
in mass found from simulations with fixed mass resolutions but initial
conditions generated using different random seeds.
  

\section{Cold Gas Fraction}
\label{sec:stats}

A critical input parameter into semi-analytical models that seek to include 
stellar as well as radiative feedback is the amount of gas in these 
first pregalactic objects that is available for star formation.  
Abel \etal (1998) estimated the fraction of gas able to cool  
by adding up all the mass within the radius of a $4\sigma$ peak (in three
different simulations) for which the temperature in radially averaged 
profiles began to decrease. However, cooling alone is not sufficient to 
determine whether the gas is available to form stars. It must also become 
dense.  We improve upon these previous estimates by measuring both the 
fraction of gas that is able to cool via molecular hydrogen cooling, $f_c$, 
and the fraction of gas that becomes both cold and dense, $f_{cd}$, in 
the clouds comprising the data sample described in Section 
\S\ref{sec:dataset}.  {\it It is this latter fraction $f_{cd}$ of cold, 
dense gas that is available to form stars in these objects.}   

\begin{figure}
\epsfxsize=7.0in
\centerline{\epsfbox{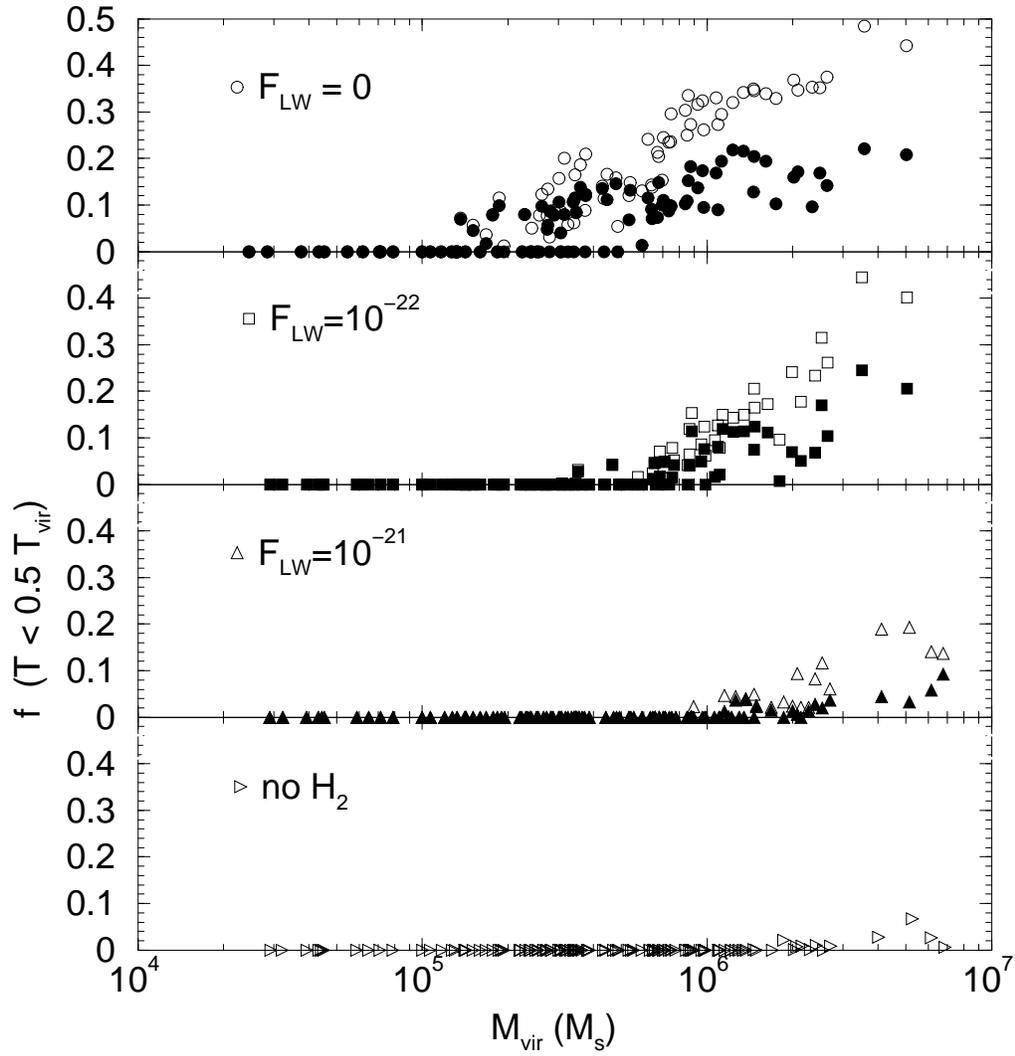}}
\caption{ Fraction of cold gas within the virial radius as functions of 
cloud mass and soft-UV background flux $F_{LW}$ in \fluxunit. Open symbols 
represent $f_c$, the fraction of gas that has cooled via \HH cooling 
($T < 0.5 T_{vir}$, $\rho > 1000 \rho_{mean}$ with $\rho_{mean}$ the mean
baryonic density of the universe). Filled symbols represent $f_{cd}$, the 
fraction of cold, dense gas ($T < 0.5 T_{vir}$, 
$\rho > 10^{19}$~$\Ms$~Mpc$^{-3}$) available for star formation.
}
\label{fig:coldfraction}
\end{figure}

In Figure \ref{fig:coldfraction} we present both these statistics as 
functions of the cloud mass $M_{vir}$ for various levels of the background
radiative flux $F_{LW}$.  We define the following:
\begin{itemize}
\item $f_c$ is the fraction of total gas within the virial radius with 
temperature $T < 0.5 T_{vir}$ and gas density $\rho > 1000 \rho_{mean}$
where $\rho_{mean}$ is the mean gas density of the universe. This is the 
amount of gas within the cloud that has been able to cool due to molecular
hydrogen cooling.
\item $f_{cd}$ is the fraction of total gas within the virial radius with
temperature $T < 0.5 T_{vir}$ and gas density $\rho > 10^{19} \MsMpc \sim
330$ \cmc .  This is the fraction of gas within the cloud that is available
for star formation.
\end{itemize}

The above criteria to determine the amount of gas in each category (cooled 
or both cold and dense) are applied on a cell by cell basis within the virial
radius of each peak in the data sample. The temperature threshold has been 
chosen to ensure that the gas is substantially cooler than the virial 
temperature of the peak as given in Equation \ref{eq:mtz}.  The density 
threshold for cold, dense gas corresponds to the gas density at which the 
baryons become important to the gravitational potential and thus to the 
subsequent evolution of the core. 
However, setting a density threshold for gas that has been able to cool 
through the \HH chemistry but may not be dense is more subtle. We must be 
careful to exclude gas that is cool only because it belongs to lower mass 
clouds that are infalling onto the more massive structure. With the high 
mass resolution in our simulations, this substructure due to merging is 
resolvable. Since there is no cooling in the peaks extracted from the 
simulation without \HH, any cold gas present must be due to infall and 
merging in the outer regions of the cloud. We confirm that this is indeed 
true in the data set without \HH cooling by looking at radial profiles for 
the cold gas fraction in several peaks. We then used this data to determine 
the density threshold for cooled gas given above that first minimizes the 
cold infalling component. As shown in the bottom panel of Figure 
\ref{fig:coldfraction}, the resultant background due to merging is 
negligible for all but the most massive peaks in the sample. The bump in 
$f_c$ in the high mass end of the bottom panel is due to the merger of the 
two most massive peaks in the box at $z  \sim 20.5$ 
seen in Figure \ref{fig:mvir_z}. Even in this extreme case the above density 
threshold keeps the infalling cold component of $f_c$ limited to a few 
percent. Note also that $f_{cd}$, the fraction of gas that can become dense 
and form stars, is zero for all peaks in the simulation without \HH, as 
expected.

Figure \ref{fig:coldfraction} represents one of the key results 
of this paper.  Both the fraction of gas $f_c$ that can cool and the 
fraction $f_{cd}$ of cold, dense gas available for star formation  
increase logarithmically with the pregalactic cloud mass over the mass 
range dominated by \HH cooling. This dependence can be summarized in a 
simple fitting formula:
\begin{equation}
    f_i = B_i \ln (M/M_{TH}) 
\end{equation} 
for $i=c$ (cooled gas) or $i=cd$ (cold, dense gas) and $M > M_{TH}$.  The 
best fit slopes $B_c$ for the fraction of gas $f_c$ that cools are $0.138 
\pm 0.006$, $0.139 \pm 0.008$, and $0.085 \pm 0.009$ for radiative 
background fluxes of $F_{LW}= 0$, $10^{-22}$, and $10^{-21}$ \fluxunit, 
respectively, where the quoted errors are standard errors of the fit.  
However, the fraction of cold, dense gas $f_{cd}$ increases much less 
rapidly with mass. Best fit slopes 
$B_{cd}$ for this cold, dense component are $0.058 \pm 0.006$, 
$0.066 \pm 0.007$, and $0.030 \pm 0.004$ for radiative background fluxes
$F_{LW}= 0$, $10^{-22}$, and $10^{-21}$ \fluxunit, respectively. The 
equality of the slopes for flux levels $J_{LW} = 0$ and $10^{-22}$ \fluxunit
is striking and probably reflects the fact that the cooling 
chemistry is the same in each case independent of radiative flux level.  The 
apparent softening of the slopes for $F_{LW} = 10^{-21}$ \fluxunit is most 
likely an artifact of the small number of independent high mass clouds in 
our simulation volume. The mass thresholds $M_{TH}$ for nonzero $f_c$ and 
$f_{cd}$ are about equal at fixed $F_{LW}$ and increase as the level of 
soft UV flux is increased.  This is a characterization of the negative 
feedback expected due to the photodissociation of \HH in the lower mass 
clouds. Our results for the mass threshold can be parameterized as:
\begin{equation}
 M_{TH}(\Ms) = 1.25 \times 10^5 + 8.7 \times 10^5 ({F_{LW} \over 10^{-21}})^{0.47}
\label{eq:mth}
\end{equation}
The presence of a soft \HH photodissociating flux at the levels considered 
here delays, but does not prevent gas in these pregalactic objects from 
becoming dense due to \HH cooling in their cores.  The fraction of gas 
available for star formation due to \HH cooling can be represented as: 
\begin {equation}
f_{cd} \approx 0.06 \ln (M/M_{TH})
\label{eq:fcd}
\end{equation}
with $M_{TH}$ given by Equation \ref{eq:mth}. 


\section{Understanding Peak Formation and Collapse}
\label{sec:profile}
 
In this section we use spherically averaged radial profiles of the 
pregalactic cloud properties to illustrate the internal dynamics of its 
collapse. We first consider the evolution with redshift of a collapsing 
pregalactic object when exposed to a fixed level of photodissociative flux.  
We then consider how the internal properties of a given cloud change at 
fixed redshift when the level of the photodissociating flux changes. 

\subsection{Evolution with redshift}

\begin{figure}
\epsfxsize=7.0in
\centerline{\epsfbox{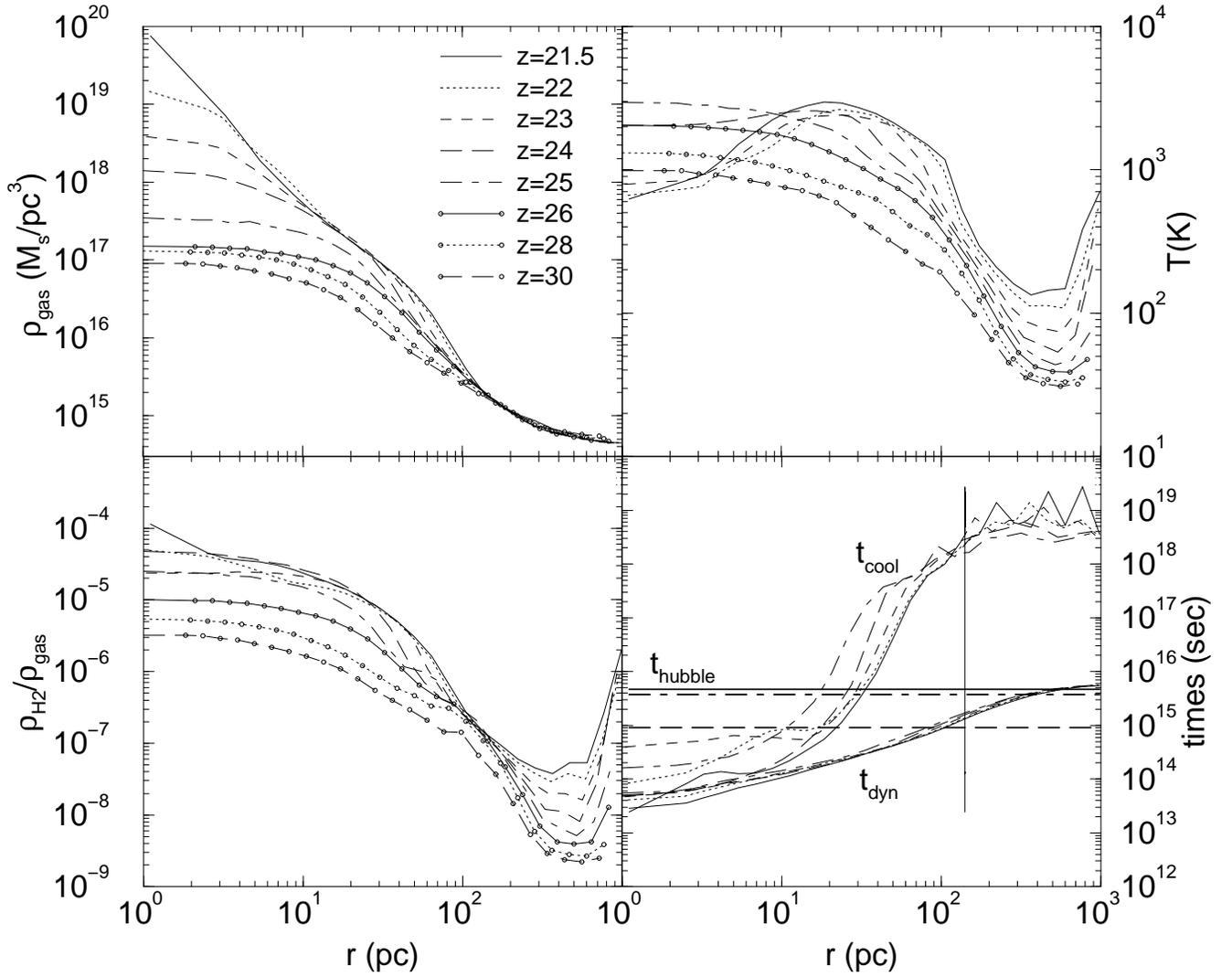}}
\caption{Evolution of the radial profiles for the first peak to collapse 
at radiative flux level $F_{LW}=10^{-21}$ \fluxunit from $z=30$ to maximal
refinement at $z=21.5$. The heavy horizontal lines in the lower right panel 
represent the Hubble time at $z=25$ (dot-dash) and at $z=21.5$ (solid), while
the heavy long-dashed line is the elapsed time from the onset of cooling until
collapse.  The vertical line marks the virial radius at $z=21.5$.}
\label{fig:P0evolve}
\end{figure}

In three of the panels of Figure \S\ref{fig:P0evolve} we show the 
evolution of the gas density, temperature, and \HH mass fraction from $z=30$ 
to maximal refinement at $z=21.5$ for the first object to collapse at a 
radiative flux level $F_{LW}=10^{-21}$ \fluxunit. 
(See Table \ref{tab:first}).  In the fourth panel (lower right) we show the 
evolution of the cooling and dynamical timescales for 
$25 \le z \le 21.5$ after cooling begins.  Starting at large radius the upper 
set of curves represent the \HH cooling time $t_{cool}$ given by 
\begin{equation}
t_{cool} = {1.5 n_g k_B T \over \Lambda n_H n_{H2}}
\label{eq:tcool}
\end{equation}
where $n_g$, $n_H$, and $n_{H2}$ are number densities for the gas,
neutral hydrogen, and \HH respectively, $k_B$ is the Boltzmann constant, $T$
is the temperature, and $\Lambda$ is the \HH cooling function. The lower set 
of curves at large radii show the dynamical timescale $t_{dyn}$ where 
\begin{equation}
t_{dyn} = \biggl({3\pi \over 16G \rho}\biggr)^{1/2}   
\label{eq:tdyn}
\end{equation}
for the same redshifts.  The solid (dot-dashed) horizontal line is the 
Hubble time
\begin{equation}
t_{Hubble} = {2 \over 3H_0} (\Omega_0 (1+z)^3)^{-1/2}
\label{eq:thub}
\end{equation}
for $z=21.5(25)$, respectively, and the heavy dashed line is the difference
between the two, i.e. the elapsed time between the onset of cooling and 
collapse.  The vertical line locates the virial radius $r_{vir} = 141$~pc
at $z=21.5$ when the structure has grown to a mass of $1.26 \times 10^6 \Ms$.
The increases in density, temperature and \HH mass fraction seen outside the 
virial radius at $\sim 1$~kpc are due to averaging over other massive 
structures forming nearby.

During the initial stages of evolution (from $30 \le z \le 25$) \HH 
abundances are too low to be an effective coolant and the cloud grows from a 
mass of $\sim 6.5 \times 10^4 \Ms$ to $5.3 \times 10^5 \Ms$ through 
accretion and merging from within its host filament. The radial profiles 
within the core are well described by constant densities and 
temperatures that increase as expected as the mass of the object increases, 
while the outer edges of the profiles steepen with decreasing $z$. The \HH 
mass fractions are also roughly constant in the core (mirroring the density 
profiles) and grow with decreasing $z$.
By $z=25$ the core temperature has risen to about $2900 \kel$, the core 
gas density has reached $\sim 3 \times 10^{17} \MsMpc \approx 10$~cm$^{-3}$, 
the \HH mass fraction is $\sim 2.3 \times 10^{-5}$ and
the core has broadened from about $10$~pc to $20$~pc. For 
$F_{LW} = 10^{-21}$ \fluxunit the photodissociation timescale 
\begin{equation}
t_{diss}(s) = k_{diss}^{-1} = 9 \times 10^{-9}/F_{LW},
\label{eq:tdiss}
\end{equation}
is $9 \times 10^{12}$~s, much shorter than other timescales in the problem. 
Thus we can understand the \HH mass fraction in the core by a simple 
equilibrium argument.  If we assume that \HH formation is dominated by the 
H$^-$ process given in Equations \ref{eq:hminusform} and \ref{eq:hminus2h2},
 the \HH equilibrium number density is given by   
\begin{equation}
n_{H2}^{equil} \approx {k_7 n_H n_e \over k_{diss}},
\label{eq:hhequil}
\end{equation}
where  $k_7 \approx 1.8 \times 10^{-18}T^{0.88}$~cm$^3$\,s$^{-1}$ over this
temperature range,
and $n_H$($n_e$) is the number density of neutral hydrogen (electrons), 
respectively. At $z=25$ Equation \ref{eq:hhequil} predicts an equilibrium 
\HH mass fraction $\chi_{H2}^{equil} \sim 2.6 \times 10^{-5}$ 
in the core that is nearly identical to the value above found in our 
simulation.
 
By $z=24$ the \HH mass fraction has reached $\sim 4 \times 10^{-5}$ 
and gas in the central region has begun to cool.  As cooling continues 
the core density steepens, evolving into a $\sim r^{-2}$ profile at
$z=21.5$ in the inner $10$~pc. The gas density reaches number densities 
$n \ge 4 \times 10^3$ exceeding the dark matter density for $r < 1$~pc. 
Notice, however, that most of the cooling occurs in the region $2<r< 20$~pc,
outside this very dense central core.  The temperature profile at 
collapse is very similar to that found by Abel, Bryan 
\& Norman (2000) showing with decreasing radius cosmic infall, shocking, and
a cooling flow. We also find for this object that the cooling time becomes 
comparable to the dynamical time over most of the cooling flow 
region just prior to collapse, and that the outer radius of the cooling region
corresponds closely to the radius at which the cooling time has dropped
below the elapsed time (heavy dashed line in Figure \ref{fig:P0evolve}) in 
agreement with the collapse criterion of Haiman, Abel \&  Rees (1999).   

Although spherically averaged radial profiles are easy to picture, it is 
important to remind ourselves that the evolution of these objects is 
fully three-dimensional and they do not collapse in 
isolation from their surroundings.  In Figure \ref{fig:image} we show
images of the log of the projected dark matter density, the baryonic
density, and the temperature at $z=21.5$ for a portion of the dynamically 
refined region in the $F_{LW}=10^{-22}$ \fluxunit simulation.  This region 
contains the two most massive structures in  
the simulation.  The image is $680$~pc (proper) on a side and is projected 
along the y-axis. The masses (virial radii) for these protogalaxies are  
$2.4 \times 10^6 \Ms$($174$~pc) for the upper cloud and $2.2 \times 10^6
\Ms$($170$~pc) for the lower cloud, while their proper center-to-center 
separation is only $\sim 300$~pc.  Thus their outer regions are already
beginning to merge.  The temperature map shows an intricate pattern of 
asymmetrical shocks with maximum temperatures around $5400 \kel$ surrounding
elongated knots of cooled gas at the clouds' centers. These cooling knots 
have temperatures of a few hundred Kelvin and are the regions of highest \HH
concentration and thus gas density.  The dark matter halos, containing 
$\sim 42,000 (37,000)$ particles for the upper (lower) structures, 
respectively, clearly show substructure near their centers due to the prior 
infall and merger of lower mass clouds that have not yet completely relaxed.
This merger activity is responsible in part for the aspherical geometry of 
the cooling knots and corresponding gas density.

\subsection{Changing the photodissociating flux}
       
\begin{figure}
\epsfxsize=2.5in
\centerline{\epsfbox{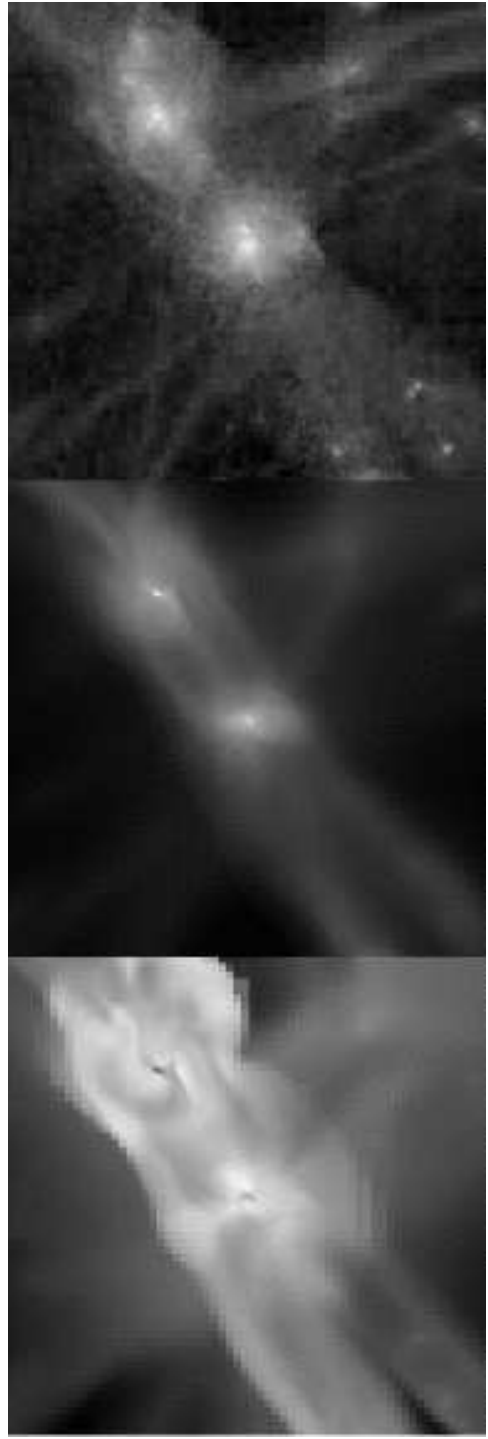}}
\caption{Two dimensional projections of the dark matter density (top), gas
density (middle), and temperature (bottom) for a region $680$~pc (proper)
on a side at $z=21.5$ in the simulation with $F_{LW}=10^{-22}$ \fluxunit.
The grayscale is logarithmic. The region contains the two most massive 
clouds, $2.4 \times 10^6 \Ms$ (upper structure) and $2.2 \times 10^6 \Ms$ 
(lower structure) formed in the simulation. 
}
\label{fig:image}
\end{figure}

\begin{figure}
\epsfxsize=5.5in
\centerline{\epsfbox{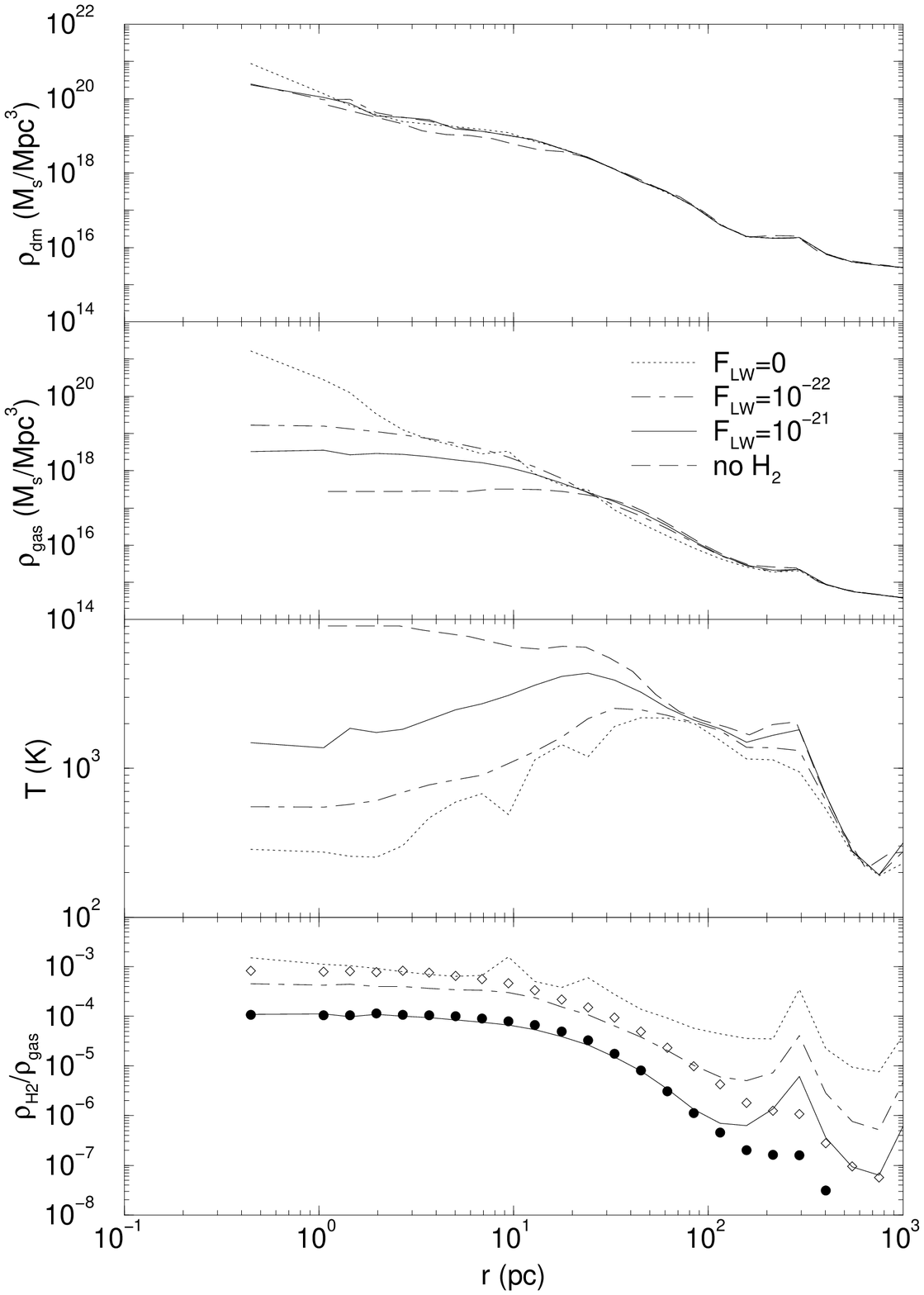}}
\caption{Radial profiles of the dark matter density, gas density, temperature
and \HH mass fraction for the lower ($M_{vir} = 2.2 \times 10^6 \Ms$, 
$r_{vir}=170$~pc) structure of Figure \protect\ref{fig:image} for various 
levels of the \HH photodissociating flux $F_{LW}$ in \fluxunit.  The 
photodissociation equilibrium values for the \HH mass fractions are also 
shown for $F_{LW}=10^{-22}$ \fluxunit (open circles) and 
$F_{LW}=10^{-21}$ \fluxunit (filled circles) for comparison. }
\label{fig:P2z21.5radprofiles}
\end{figure}

In Figure \ref{fig:P2z21.5radprofiles} we compare the spherically averaged 
radial profiles of dark matter density, gas density, temperature and \HH
mass fraction at fixed redshift ($z=21.5$) but with varying levels of \HH
photodissociating flux for the $2.2 \times 10^6 \Ms$ (lower) 
structure of Figure \ref{fig:image}. The cloud has only been able to 
maximally refine in the simulation with $F_{LW}=0$, although it is close to 
collapse in the $F_{LW}=10^{-22}$ case also.  This is in spite of  
the fact that the mass of this object is above the masses of the 
first objects to collapse in both the $F_{LW}=10^{-22}$ and $10^{-21}$ 
\fluxunit simulations (See Table \ref{tab:first}).  Probably this is due
to the strong effect of tidal interactions and high level of merging 
activity in and surrounding this object.  The features located at $300$~pc 
and $1$~kpc in the profiles reflect the presence of the upper, 
$2.4 \times 10^6 \Ms$ protogalaxy of Figure \ref{fig:image} and the 
$1.26 \times 10^6 \Ms$ structure we analyzed earlier in Figure
\ref{fig:P0evolve}, respectively.  The dark matter density profiles are 
essentially the same for all of the simulations, independent of \HH fraction 
or radiative flux.  This is to be expected as long as the gravitational 
potential is dominated by the dark matter.  The dark matter density 
only steepens away from this common profile in the inner $\sim 2$~pc of the 
$F_{LW}=0$ case when the gas density equals and then exceeds the dark matter 
density.  For the $F_{LW}=0$ case the gas density steepens
in the inner several parsecs with gas number densities reaching 
$\sim 10^4$, characteristic of the formation of a dense collapsed core. The 
cooling region extends inward from about $40$~pc until the temperature drops 
to the minimum value possible under \HH cooling, $\sim 200 \kel$, at around
$3$~pc. The \HH mass fraction has saturated at $\sim 10^{-3}$. The sharp 
drops in temperature that correspond to enhancements in the 
\HH mass fraction at $9$ and $25$~pc are the result of averaging over 
infalling clouds that were able to form \HH in their cores and cool 
before merging with this more massive object.  As the radiative flux level 
is increased, the negative feedback becomes 
apparent. The maximum gas density drops in the central region and forms a 
homogeneous core whose core radius also increases as the flux is increased. 
The radius of the cooling region moves inward and cooling is suppressed 
due to the decreased amount of \HH coolant.  The cloud is clearly in 
photodissociation equilibrium for the high flux ($F_{LW}=10^{-21}$) case with
the \HH fraction given by Equation \ref{eq:hhequil}.  For the lower flux 
case, the \HH fraction is somewhat smaller than the photodissociation 
equilibrium value in the inner $10$~pc due to the influence of
recombination on the formation of H$^-$ ions and thus the \HH formation rate.
We also see, as expected, that there is no cooling at all for the case 
in which \HH is absent.   

\begin{figure}
\epsfxsize=5.5in
\centerline{\epsfbox{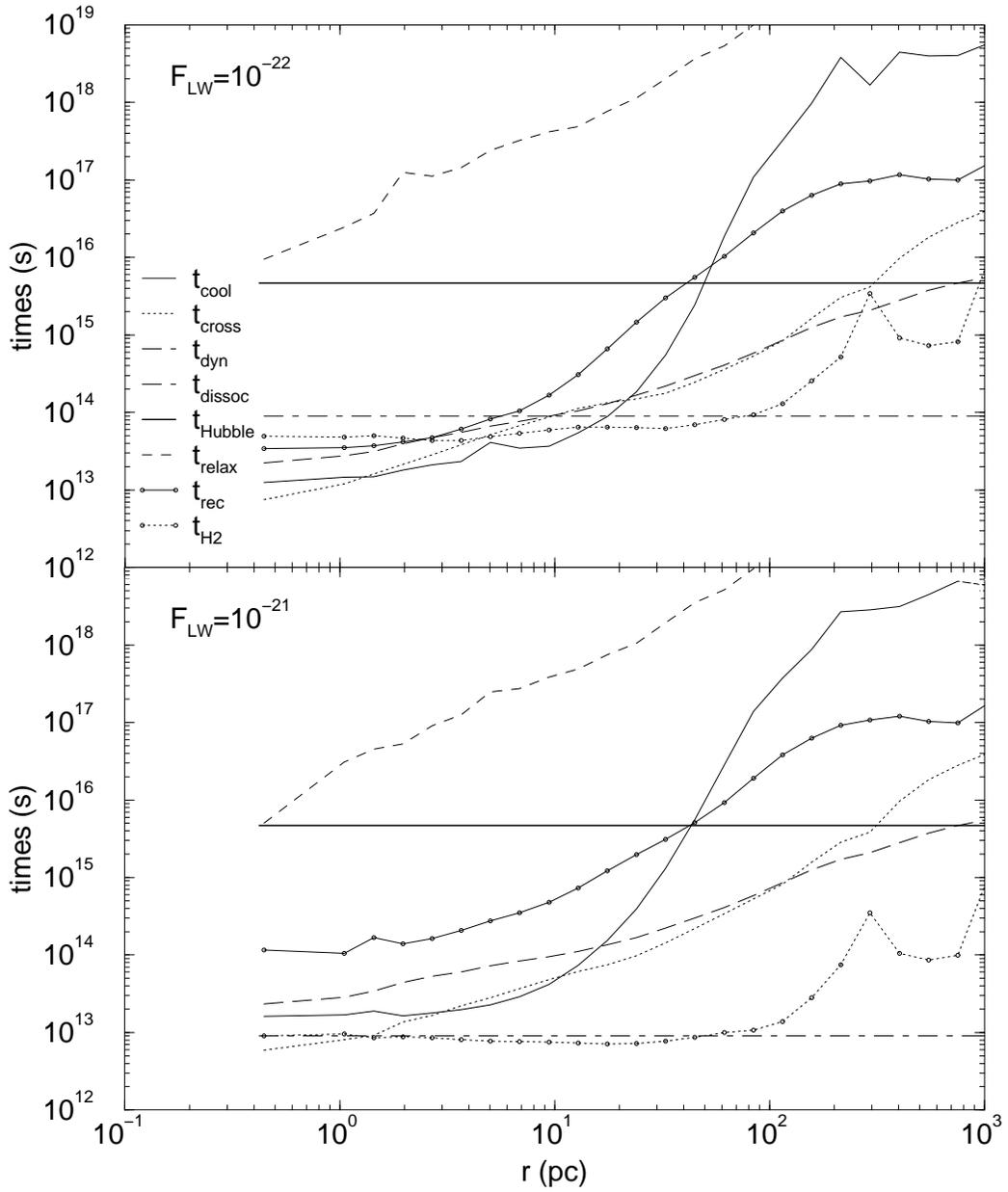}}
\caption{Radial profiles of timescales relevant to cooling and collapse at 
$z=21.5$ for the $2.2 \times 10^6 \Ms$ structure of Figure 
\protect\ref{fig:image} at redshift $z=21.5$ and background soft-UV flux 
levels $F_{LW}=10^{-22}$ \fluxunit (top) and $10^{-21}$ \fluxunit (bottom).
}
\label{fig:P2times}
\end{figure}

We can gain a better understanding of the above radial profiles and their 
scaling properties by comparing the relevant physical 
timescales in the problem.  These times are shown
for both the $F_{LW}=10^{-22}$ case (upper panel) and the 
$F_{LW}=10^{-21}$ case (lower panel) in Figure \ref{fig:P2times}.
In addition to profiles for the Hubble, dynamical, photodissociation, and 
cooling times defined earlier, we include profiles for the sound 
crossing time $t_{cross}$, the recombination time $t_{rec}$ given by 
\begin{equation}
t_{rec} = 7.7 \times 10^9 T^{0.64}/n_e ,
\label{trec}
\end{equation}
and the \HH formation time $t_{H2}$ given by  
\begin{equation}
t_{H2} = {n_{H2} \over k_7 n_H n_e}.
\label{eq:tH2}
\end{equation}
All times are in seconds. We also show that the two-body relaxation time 
$t_{relax}$ (Equation 8-71 in Binney \& Tremaine 1987) for the dark matter 
particles is long compared to all other timescales confirming that two-body 
numerical effects are unimportant.

For both cases the photodissociation and \HH formation times are much
shorter than the Hubble time everywhere within the virial radius.  In
the high flux case they are also much shorter than the recombination
time so that photodissociation equilibrium ($t_{diss}=t_{H2}$) is
established throughout the cloud (see equation \ref{eq:hhequil}).
Even so, enough \HH forms for the cooling time to drop below the
Hubble time in the inner $40$~pc --- core collapse is not far off.
The low flux case is similar in the outer parts of the cloud: the
recombination time is long and the \HH fraction is governed by its
equilibrium value.  However, within $10$~pc recombination becomes
faster than photodissociation.  This reduces the amount of $H^-$
allowed to form and suppresses \HH formation below the
photodissociation equilibrium value.  However, enough \HH has formed
that cooling and collapse proceed faster than photodissociation
($t_{dyn} < t_{diss}$ and $t_{cool} < t_{diss}$) at these small radii.
We also find here --- as in Abel, Bryan, \& Norman (2000) --- that
$t_{cross} < t_{dyn}$ in the inner regions of the cloud so that the
collapse proceeds hydrostatically.


\section{Discussion}
\label{sec:discussion}

Now that we have detailed examples of the collapse properties of clouds in 
our simulations, we are better able to assess the importance of neglected
processes in our simulations.  In this section we discuss in more detail 
why the neglect of self-shielding and H$^-$ photodetachment should not 
significantly change our results.  We then use a simple analytical model,
similar to that by Tegmark \etal (1997), to illustrate the essential physics
of the collapse. 

\subsection{Neglected processes}
\label{sec:neglect}
 
We first return to the issue of $H_2$ self-shielding.  Although this
effect could, in principle, be important for our objects, we argue
here that its actual impact is slight.  The maximum column densities
of \HH we encounter in our simulations are around $10^{17}$ cm$^{-2}$,
while a column density of $5 \times 10^{14}$ cm$^{-2}$ is sufficient
to start blocking the soft-UV flux.  So, at first glance, it appears
that self-shielding should be important; however, this value has been
derived for a static gas distribution.  The Lyman-Werner bands consist
of a large number of individual lines whose width is dominated (in our
case) by Doppler broadening.

There are substantial velocity gradients within all of our simulated
objects, even those which appear to be relaxed.  These take the form
not just of radial inflow, but also large-scale disordered velocity
fields (i.e. turbulence).  Typical {\it rms} line-of-sight velocities
are around 3-4 km/s.  These values are larger than the $\sim$2 km/s
line widths due to thermal Doppler broadening.  This means that even
if a given parcel of gas is behind a large column of molecular
hydrogen, it may still be photo-dissociated if it has a velocity shift
relative to the blanketing material.  There is very little continuum
opacity.  

It is a difficult radiative transfer calculation to determine the
precise effect of these velocity shifts; however we may consider two
limiting cases.  The first is the previously mentioned static case in
which velocities are negligible compared to the thermal Doppler widths
($v/b \ll 1$) and an $H_2$ column density of $5 \times 10^{14}$
cm$^{-2}$ is sufficient to start affecting the photo-dissociation
rate.  In the other extreme, in which $v/b \gg 1$, the rate is not
significantly reduced until a substantial fraction of the 11-13 eV
range is blocked by the Lyman-Werner lines.  Since this does not occur
until the damping wings of the lines become important, it requires a
much larger molecular hydrogen column density, of order $10^{22}$
cm$^{-2}$ (Draine \& Bertoldi 1996; Glover \& Brand 2000).  In the
simulations, $v/b > 1$ so we are likely to be closer to the second
regime than the first.  Even if it were marginally important, there
are three further reasons to think that self-shielding is unimportant.
First, these large column densities occur at late times, once the core
has already collapsed.  For most of the evolution the objects are
clearly in the optically thin regime.  Second, the maximum column
density quoted above is to the center of the object, but the cooling
front lies outside of the $H_2$ core, and so beyond most of the
covering material.  Finally, once a sufficient amount of $H_2$ has
formed to produce a substantial blanketing column, the cooling time
has dropped below the photo-dissociation time, so that
photo-dissociation is no longer important (and thus even if the
soft-UV flux is attenuated, the subsequent evolution of the gas would
be unchanged).  So while the results described earlier are properly
lower limits to the cooled fraction because of our neglect of
self-shielding, a full radiative transfer calculation should produce
nearly identical results.

Let us also reexamine whether H$^-$ photo-detachment (which we
neglect) could significantly affect the results.
In order to bracket the size of this effect we consider two spectral
shapes for the radiation field, $J \sim \nu^{\alpha}$ with $\alpha =
1$ and $-1$. The softer spectrum, $\alpha =1$, is the spectral slope
expected in this energy range from a typical OB star. The steeper
slope, $\alpha = -1$, is the type of spectral shape one might expect
for a background radiation field built up from a number of
sources.  
The cross section (in cm$^2$) for H$^-$ photo-detachment is given by 
(Kang \& Shapiro 1987, Abel \etal 1997),
\begin{equation}
\sigma_{detach} = 7.928 \times 10^5 
     (\nu - \nu_{th})^{3/2}{1 \over \nu^3}  \quad 
     {\rm for}\, \nu > \nu_{th}
\end{equation}
where $h_P$ is Planck's constant and $h_P \nu_{th} = 0.755$ eV.
The H$^{-}$ photo-detachment rate coefficient $k_{detach}$ 
can then be calculated
\begin{equation}
k_{detach} = \int_{\nu_{th}}^{\nu_{H}} {4\pi J(\nu) \over h_P\nu} \sigma_{detach}(\nu)\, d\nu
\end{equation}
with $h_P\nu_{H}=13.6$~eV.    
For the two spectral shapes given above we find $k_{detach}=1.9
\times 10^{-12} F_{LW}$~s$^{-1}$ for $\alpha = 1$ and $6.4 \times
10^{-11}F_{LW}$~s$^{-1}$ for $\alpha = -1$ where $F_{LW}$ is the radiation 
flux in \fluxunit. 

The dominant processes other than photodetachment
that determine the abundance of H$^-$ are shown in Equations 
\ref{eq:hminusform} and \ref{eq:hminus2h2} with rate coefficients $k_7$ and 
$k_8$, respectively. Furthermore the processes proceed sufficiently rapidly
compared to the \HH chemistry that $n_{H-}$ is well approximated by its 
equilibrium value (Abel \etal 1997) such that (including photo-detachment):
\begin{equation}
n_{H-} \approx {k_7 n_H n_e \over k_8 n_H + k_{detach}}.
\end{equation}
We see that H$^-$ photo-detachment can be neglected as long as
$k_{detach} \ll k_8 n_H$. This leads to a condition on the hydrogen
number density $n_H \gg k_{detach}/k_8$ that can be checked against
the conditions found within the collapsing structures.  For gas
temperatures found in our simulations, $k_8 \approx 1.43 \times
10^{-9}$ cm$^3$s$^{-1}$ (Abel \etal 1997). Thus photo-detachment is
unimportant provided that $n_H \gg 0.045(F_{LW}/10^{-21})$ for the
steeper $\alpha = -1$ spectrum or, a much weaker constraint, that $n_H
\gg 1.4 \times 10^{-3}(F_{LW}/10^{-21})$ for the softer $\alpha = 1$
spectrum.  Let us check these constraints against the hydrogen number
densities found within the first peak to collapse at our highest flux
level ($F_{LW} = 10^{-21}$~\fluxunit) shown in Figure
\ref{fig:P0evolve}.  At $z=25$, before cooling commences, hydrogen
number densities in the core region where \HH will form already exceed
$n_H \sim 7$.  This number density satisfies the constraint for the
steepest spectrum ($\alpha =-1$) by more than two orders of magnitude
so that the neglect of H$^{-}$ photo-detachment does not significantly
affect our results.

\subsection{An Analytic Estimate of the Collapse Threshold}
\label{sec:simple} 

Analytical arguments often provide qualitative insight into the 
key physical principles at work in a complex process. Since we have seen 
that the \HH fraction is close to photodissociation equilibrium throughout
our objects, we use these equilibrium abundances to develop analytical 
estimates of the critical \HH fraction and mass thresholds for collapse. 
We adopt the simplified collapse criterion, similar to that of Tegmark \etal
(1997), that an object will collapse if the cooling time $t_{cool}$, given
by Equation \ref{eq:tcool}, is less than the Hubble time $t_{Hubble}$, 
defined in Equation \ref{eq:thub}.  Thus  
\begin{equation}
{1.5 n_g k_B T \over \Lambda n_H n_{H2}} \le {2 \over 3H_0} (\Omega_0 (1+z)^3)^{-1/2} 
\end{equation}
The equality gives the critical number density of \HH necessary for collapse.
Solving for $n_{H2}^{crit}$ we obtain 
\begin{equation}
n_{H2}^{crit} = 2.25 {n_g k_B T \over n_H \Lambda} H_0 \Omega_0^{1/2} (1+z)^{3/2}
\label{eq:ncrit}
\end{equation} 
In this equation, $\Lambda$ is, in our case, the Lepp \& Shull 
(1983) cooling function. In the low density limit ($n_{H2} \ll 10^4$ \cmc) 
for temperatures $300 \kel \le T \le 5000 \kel$ of interest here it is well 
approximated by $\Lambda \approx 6.0 \times 10^{-11}T k_B e^{-730/T}$ such 
that 
\begin{equation}
n_{H2}^{crit} \approx 1.3 \times 10^{-7}he^{730/T}\Omega_0^{1/2}(1+z)^{3/2}
\label{eq:ncrit2}
\end{equation}
If we now equate this critical abundance of \HH required to satisfy our 
collapse criterion to the photodissociation equilibrium
abundance from Equation \ref{eq:hhequil} and solve for $F_{LW}$,
we obtain an expression for the critical flux that just allows
a cloud with virial temperature $T$ to collapse.
\begin{equation}
\biggl ({F_{LW}^{crit} \over 10^{-21}}\biggr ) \approx {126 n_H^2 x_e T^{0.88} e^{-730/T} \over h \Omega_0^{1/2} (1+z)^{3/2}}.  
\label{eq:fluxcrit}
\end{equation}
where the electron fraction $x_e = n_e/n_H$.

To complete the above relationship we need to approximate the electron
fraction and the gas number density in the cloud core.  If the
collapse occurs well within one recombination time, the electron
fraction, $x_e$, is expected to be close to its residual value $x_e =
1.2 \times 10^{-5} \Omega_0^{1/2}/(\Omega_b h)$.  The average baryon
number density in the universe is given by $n_B \sim 9.3 \times
10^{-6}(1+z)^3 \Omega_b h^2$~\cmc. The gas temperature after thermal
decoupling from the CMB at $z \sim 200$ but before reionization $z >
7$ drops adiabatically with $T_{IGM} \sim 0.0135(1+z)^2 \kel$. If we
assume that the baryons in the central regions of dark matter halos
were never shocked, we find the maximum density that can be reached by
adiabatic evolution is $n_{max} =
n_B(T_{vir}/T_{IGM})^{1/(\gamma-1)}$. Here $T_{vir}$ denotes the
virial temperature of the dark matter halo and $\gamma = 5/3$ is the
ratio of specific heats for an ideal gas.  This gives a collapse
redshift independent maximal density as a function of virial
temperature for adiabatic evolution of the cloud as $n_{max} \sim 187
\Omega_b h^2 (T/1000)^{3/2}$ \cmc. We expect adiabatic evolution to be
a good approximation for low mass clouds, but caution that for larger
mass structures merging and shock heating become important.  Inserting
the expressions for $n_{max}$ and $x_e$ into Equation
\ref{eq:fluxcrit} we obtain:
\begin{equation}
\biggl ({F_{LW}^{crit} \over 10^{-21}}\biggr ) \approx 223 \Omega_b h^2 \biggl ({T \over 1000}\biggr )^{3.88}e^{-730/T} \biggl ( {1 + z \over 20} \biggr )^{-3/2}
\label{eq:fluxcrit2}
\end{equation}

The structures that are first able to collapse in our simulations all have 
virial temperatures between $700 \kel$ and $3000 \kel$ where the exponential
in Equation \ref{eq:fluxcrit2} can be approximated by a power law,
$e^{-730/T} \approx 0.48(T/1000)^{0.7}$. Inserting this approximation into
Equation \ref{eq:fluxcrit2} and inverting, we obtain an analytic 
approximation for the critical virial temperature for collapse at a given 
soft-UV flux level.
\begin{equation}
\biggl ({T_{vir}^{crit} \over 1000}\biggr ) \approx 0.36 \Biggl( (\Omega_b h^2)^{-1}\biggl ({ F_{LW} \over 10^{-21}} \biggr ) \biggl ( { 1+z \over 20}\biggr )^{3/2}\Biggr )^{0.22} 
\label{eq:Tcrit}
\end{equation}
For $F_{LW}=10^{-21}$ \fluxunit at $z=25$ in our cosmological model 
($h=0.65$, $\Omega_b=0.05$) this gives a $T_{vir}^{crit} \sim 920 \kel$, 
somewhat lower than but comparable to the $T_{vir} \sim 1470 \kel$ found at 
that redshift for the first peak to reach maximal refinement at that flux 
level in our simulation (See Figure \ref{fig:P0evolve}).  The 
disagreement in detail is not surprising since the approximation invokes a 
simplified collapse criterion and ignores the nonadiabatic evolution of the 
collapse. However, the analytical argument does serve to ellucidate the 
essential physics at work in the collapse.


\section{Summary}
\label{sec:summary}

In this paper, we have used high-resolution numerical simulations to
investigate the effect of radiative feedback on the formation of
$10^5-10^7 M_{\odot}$ clouds.  This range of objects is important
because they are large enough to form molecular hydrogen, but too
small to cool by hydrogen line cooling.  Therefore they are
susceptible to the destruction of molecular hydrogen from soft-UV flux
in the 11-13 eV range (the Lyman-Werner bands).  This radiation comes
from the very first generation of stars and so its amplitude is
unclear; we test four cases, ranging from zero flux to complete \HH
destruction.  The initial conditions are drawn from a $\Lambda$CDM
cosmological model, and the simulations evolve the non-equilibrium
rate equations for twelve species of hydrogen and helium.  Our main
conclusions may be summarized as follows.
\begin{itemize}
\item A soft-UV flux does delay cooling and star formation in the
cores of clouds in this mass range.  An increasing amount of radiation
delays collapse until later redshifts when larger objects have
collapsed.  However, even without any radiative feedback, objects with
masses less than $\sim 1.4 \times 10^5 M_{\odot}$ (at $z \sim 30$) do
not cool efficiently due to their low \HH fraction.  We provide a fit
to the mass threshold for collapse, given the mean flux in the
Lyman-Werner bands (equation \ref{eq:mth}).
\item The amount of cold, dense gas which is available for star
formation in the cores of the collapsed objects depends primarily on
two numbers, with relatively little scatter: the flux of soft UV
radiation and the mass of the cloud.  We show the cold, dense fraction
increases logarithmically with the mass of the halo and we provide a
fit for this relation as well (equation \ref{eq:fcd}).
\item The evolution of these objects is relatively straightforward.
With a negligible soft-UV radiation field, the formation of \HH
continues until it is suppressed by the removal of electrons from the
core due to recombination.  However, by the time this occurs,
generally enough \HH has formed to allow core collapse.  When
radiative feedback is important ($F_{LW} \gtrsim 5 \times 10^{-24}$
\fluxunit), the destruction and formation processes are sufficiently
fast so that the \HH fraction is in equilibrium
(equation \ref{eq:hhequil}) and recombination plays little role until
after core collapse begins.  This allows the derivation of an
analytic estimate for the collapse threshold which comes close to
matching the full simulation results.
\end{itemize}

Finally, there are a number of other results which deserve
further investigation.  One of these is that the collapse redshift
does not seem to depend on the angular momentum of the halo (see
section~\S\ref{sec:dataset}).  Another is that we do not appear to
witness fragmentation in these objects.  The few times in which we
clearly see more than one collapsed core in a halo appears to be
due to the merging of a number of halos, each of which already
contained a core.

We also examined a number of process which we omitted from the
simulation, including \HH self-shielding and $H^-$ photo-detachment,
none of which appears to be important for this calculation.  These
results were derived within the context of a currently popular
$\Lambda$-dominated cosmological model.  However, at these high
redshifts, the effective $\Omega$ should be close to one, so the most
important parameter is the amplitude of the power spectrum on these
scales.  Relatively small changes to the amplitude (such as would
arise from the current uncertainty in $\sigma_8$) are unlikely to have
a major impact on the results presented here, although it should
always be kept in mind that we are extrapolating the power spectrum to
scales significantly smaller than constrained by observations.



\acknowledgements

This work is supported in part by  NSF grant AST-9803137 under the auspices 
of the Grand Challenge Cosmology Consortium.  NASA also supported 
this work through Hubble Fellowship grant HF-0110401-98A from the Space 
Telescope Science Institute, which is operated by the Association of 
Universities for Research in Astronomy, Inc. under NASA contract 
NAS5-26555.  The computations used the SGI Origin2000 at the National
Center for Supercomputing Applications.  We acknowledge useful
discussions with Zoltan Haiman.



\newpage

\begin{deluxetable}{cccccccccc}
\footnotesize
\tablecaption{Properties of the first object to reach maximal refinement for 
various levels of soft-UV background flux $F_{LW}$.
\label{tab:first}}
\tablewidth{0pt}
\tablehead{
\colhead{$F_{LW}$(erg/s/cm$^2$/Hz)} & 
\colhead{$z$} & 
\colhead{$M_{total}$} & 
\colhead{$T_{virial}$(K)} & 
\colhead{$r_{virial}$(pc)} & 
\colhead{$M_{dm}$} & 
\colhead{$M_{gas}$} & 
\colhead{$M_{cd}$} & 
\colhead{$f_{cd}$} & 
\colhead{$\lambda_{dm}$}
}
\startdata
$0$  & $30.2$ & $1.37 \times 10^5$ & $713$   & $49$  & $1.28 \times 10^5$  & 
$8800$ & $616$ & $0.07$ & $0.023$ \\
$10^{-22}$ & $24.4$ & $8.83 \times 10^5$ & $2016$ & $111$ & 
$8.02 \times 10^5$ &
$8.08 \times 10^4$ & $1734$ & $0.021$ & $0.107$ \\
$10^{-21}$  & $21.5$ & $1.26 \times 10^6$ & $2262$ & $141$ & 
$1.15 \times 10^6$ & $1.15 \times 10^5$ & $4180$ & $0.037$ & $0.042$ \\
\enddata
\tablecomments{
$z$ is the redshift of maximal refinement. $T_{vir}$ is the virial
temperature given by Equation \protect\ref{eq:mtz}. $M_{total}$,
$M_{dm}$, and $M_{gas}$ are the total, dark matter, and gas masses,
respectively, in $M_\odot$ within the virial radius $r_{vir}$.
$M_{cd}$ ($M_\odot$) is the total amount of cold, dense gas ($T
< 0.5 T_{vir}$, $\rho > 10^{19} \Ms$~Mpc$^{-3}$) within the virial
radius, $f_{cd}$ is the corresponding cold, dense gas fraction, and
$\lambda_{dm}$ is the dark matter halo spin parameter.
}
\end{deluxetable}

\end{document}